\documentclass{article}
\usepackage[utf8]{inputenc}
\usepackage{authblk}
\usepackage{hyperref}
\usepackage{graphicx}
\usepackage{xcolor}
\usepackage{amsmath}
\usepackage{adjustbox}
\usepackage{array, booktabs, makecell}
\usepackage{url}
\usepackage{geometry}
\usepackage{fancyhdr}
\usepackage{algorithmic}
\usepackage{algorithm}
\title{Semi-supervised time series classification method for quantum computing}

\author[1,2]{Sheir Yarkoni\footnote{These authors contributed equally to this work.}\footnote{Corresponding author: \underline{sheir.yarkoni@volkswagen.de}}$^{,}$}
\author[1]{Andrii Kleshchonok$^{*,}$}
\author[1]{Yury Dzerin}
\author[2, 3]{Florian Neukart}
\author[1]{Marc Hilbert}
\affil[1]{Volkswagen Data:Lab, Munich, Germany}
\affil[2]{LIACS, Leiden University, The Netherlands}
\affil[3]{Volkswagen Group of America, San Francisco, USA}
\date{}

\makeatletter
\newcommand{\HEADER}[1]{\ALC@it\underline{\textsc{#1}}\begin{ALC@g}}
\newcommand{\ENDHEADER}{\end{ALC@g}}
\makeatother

\begin{document}

\maketitle

\begin{abstract}
    In this paper we develop methods to solve two problems related to time series (TS) analysis using quantum computing: reconstruction and classification. We formulate the task of reconstructing a given TS from a training set of data as an unconstrained binary optimization (QUBO) problem, which can be solved by both quantum annealers and gate-model quantum processors. We accomplish this by discretizing the TS and converting the reconstruction to a set cover problem, allowing us to perform a one-versus-all method of reconstruction. Using the solution to the reconstruction problem, we show how to extend this method to perform semi-supervised classification of TS data. We present results indicating our method is competitive with current semi- and unsupervised classification techniques, but using less data than classical techniques. 

\end{abstract}

\section{Introduction}
The field of quantum computing has experienced rapid growth in recent years, both in the number of quantum computing hardware providers and their respective processors' computing power. Companies such as D-Wave Systems, Rigetti, and IBM offer access to their quantum processors, and their use in proof-of-concept demonstrations has been widely discussed in literature. Quantum processing units (QPUs) have been used to solve a wide variety of problems such as traffic flow~\cite{trafficflow}, logistics and scheduling~\cite{jobshop, flightscheduling}, quantum simulation~\cite{vwchemistry, McCaskey2019, Grimsley2019}, and more~\cite{Venturelli2019, advertising}. Notably, a recent study by Google showed how their QPU can perform the task of sampling from random quantum circuits faster than state-of-the-art classical software~\cite{quantumsupremacy}, ushering a new era in the field of quantum computing. These applications use so-called noisy intermediate scale quantum (NISQ~\cite{nisq}) processors to solve various forms of optimization and sampling problems. Most commonly, the problem is formulated as a quadratic unconstrained binary optimization (QUBO) problem, or its equivalent form of an Ising Hamiltonian. The former uses a basis of binary $\{0, 1\}$ variables, and the latter makes use of spin variables $\{-1, 1\}$. Both can be solved using existing quantum computing hardware. \\ 

The QPUs provided by D-Wave Systems use a quantum annealing algorithm that implements a transverse-field Ising Hamiltonian~\cite{manufacturedspins}. This quantum protocol prepares an initial Hamiltonian with a simple ground state, and transitions to a Hamiltonian whose ground state is difficult to find. This is referred to as Adiabatic Quantum Computation (AQC)~\cite{aqc}, and under open quantum system conditions as quantum annealing~\cite{nishimori}. Because AQC has been shown to be polynomially equivalent to gate-based quantum computation~\cite{aharonov2008adiabatic}, and the Ising spin-glass has been shown to be NP-hard to minimize~\cite{barahona}, AQC (and quantum annealing) has the potential to significantly impact the fields of optimization, machine learning, and operations research. Equivalently, with gate-model QPUs such as those produced by Google, IBM, and Rigetti, the quantum approximate optimization algorithm (QAOA) is used to solve such Ising Hamiltonians. This algorithm also attempts to minimize a target Ising Hamiltonian by alternating between a driver and mixer Hamiltonian, until the sampling procedure converges to the target state population. The derivation and details of the QAOA algorithm are beyond the scope of this paper, and are discussed in detail in~\cite{qaoa}. \\

At the end of a quantum annealing run, or a QAOA circuit execution, the measurements are a projection along the $z$-component of the qubits' spins, resulting in a sequence of classical bit strings. These states can be interpreted as approximations to finite-temperature Boltzmann states from the classical spin-glass Ising Hamiltonian~\cite{temperature, qaoaboltzmann}:

\begin{equation}
        H(s) = \sum_i h_i s_i + \sum_{i<j} J_{ij} s_i s_j.
\end{equation}

Therefore, the task of programming a quantum annealer or QAOA circuit involves finding a suitable Ising Hamiltonian representation for the optimization task. In this paper, we motivate classification of time series data based on extracting features which exist within the data, and use combinatorial optimization techniques to match and reconstruct data with other time series. We start by reducing the dimensionality of the TS data and encode it as a string. We introduce a pulling procedure, comparing the encoded strings to form a collection of sets, where common features between the strings are extracted. All extracted common features are pooled together, which we can then use to construct new TS and compare between existing ones. We perform these tasks by using the pulled features as elements of the universe in the set cover problem, which has a known QUBO/Ising formulation and can be solved using quantum computers. By reformulating the critical task in our clustering algorithm as a set cover problem, we introduce two novel ideas to quantum clustering algorithms: (1) we avoid representing single vectors with polynomial numbers of qubits, instead representing the features within the data as the qubits, and (2) we perform the clustering task by transferring the core concepts of clustering (and reconstruction) to the quantum algorithm for set cover, as opposed to a direct translation of a distance-based minimization procedure. \\

The rest of this paper is organized as follows. Section~\ref{sec:previous} provides a short overview of existing methods for both classical and quantum clustering. Section~\ref{sec:problem_formulation} motivates the task of TS reconstruction, explains the methods used to discretize the data, and how to convert the discretized data to the set cover problem and its representative QUBO. Section~\ref{sec:clustering} shows how to extend the reconstruction method to classify the TS data. Section~\ref{sec:results} outlines the experiments performed to test the developed method using various open-source data sets, and conclusions are presented in Section~\ref{sec:discussion}.

\section{Previous works}
\label{sec:previous}
Quantum computing-based approaches which exist in literature involve fundamentally different approaches than that introduced in this work; we provide a brief overview of some key methods and algorithms related to quantum clustering. Assuming the existence of error-corrected quantum processors (and the existence of quantum RAM), it has been shown that quantum computers could perform $k$-means clustering exponentially faster than their classical counterparts~\cite{Lloyd2013QuantumAF}. Other works have also shown how to reformulate parts of classical clustering algorithms as quantum subroutines that can be executed on error-corrected gate-model QPUs~\cite{quantum_TS1,quantum_clustering2,quantum_clustering3,quantum_clustering4}. In quantum annealing, a similar approach has been shown in which the objective function of the clustering task (minimizing distance metrics between high-dimensional vectors) has been directly translated to a QUBO, with each vector's possible assignment represented via one-hot encoding to physical qubits~\cite{booz, quantum_clust}. \\

Classical time series (TS) analysis is considered to be a challenging task due to high number of dimensions involved resulting in the Curse of Dimensionality phenomenon. A series of works address the question of efficient dimensionality reduction~\cite{HOTSAX,symbol2,symbol3,senin2018grammarviz,schafer2012sfa,Patel2002MiningMI,PAA_paper,1021475}, explaining the trade-off between information loss and search space size. Main results presented in this manuscript are obtained with Symbolic Fourier Approximation (SFA) method~\cite{schafer2012sfa} due to its pruning power, noise-robustness and scalability. SFA represents each real-valued TS in a frequency domain by a symbolic string using the discrete Fourier transform. These transformed TS can then be used by classical string-based similarity algorithms such as phonetic distance based, Levenshtein, Hamming, Jaro, Jaro-Winkler measures, and more~\cite{gomaa2013survey}.  \\

Classical TS clustering techniques can be split into the following categories: model-based, feature-based, shape-based and their combinations \cite{TS_clust4}. In the model-based approach the TS is encoded and fit by parametric models and clustering is applied to these extracted parameters~\cite{TS_clust2}. In feature-based methods, the features of TS, like Fourier components, periodicity, trend, number of peaks, variance, etc., are extracted and later clustered by conventional algorithms~\cite{Hautamaki2008TimeseriesCB, CHRIST201872}. Shape-based approaches refer to comparing shapes of TS directly and matching them according to specifically chosen metrics. A typical example for this approach is Dynamic Time Warping (DTW)~\cite{DTW_original}, which has been shown to outperform Euclidean metrics \cite{Chu2002IterativeDD, Vlachos2002DiscoveringSM}. DTW-based classical methods are used to evaluate the accuracy of our approach in Section~\ref{sec:results}. For more details on classical approaches to TS clustering, we refer the reader to~\cite{TS_clust1,TS_clust3,TS_clust4}.\\

\section{Time Series Reconstruction: Problem Formulation}
\label{sec:problem_formulation}
Clustering techniques generally require specific data representation, similarity measure definitions, and clustering algorithm selection. Similarly, in our quantum computing based approach, we represent the TS data as encoded strings from which we formulate semi-supervised clustering and optimal reconstruction as a set cover problem, and provide metrics based on solutions to the set cover problem. While different than classical approaches~\cite{reconstr2,reconstr1, reconstr3, reconst4}, we preserve the computational complexity of the problem, while introducing a method that is based on latent features within the data. \\  \newpage

In order to reconstruct given time series data, we start by discretizing the data, and comparing the encoded strings to generate the elements of our universe to form the set cover. This pulling technique is crucial to allow feature-wise comparison of the data, as well as arbitrary reconstruction of TS using existing (or training) data. We use existing techniques for discretization, and explain the pulling procedure in detail. We then show how to use this data to construct the set cover problem for quantum optimization.

\subsection{Discretization and Pulling Technique}
\label{sec:pulling}
\begin{figure}[ht]
\centering
        \includegraphics[totalheight=4.8cm]{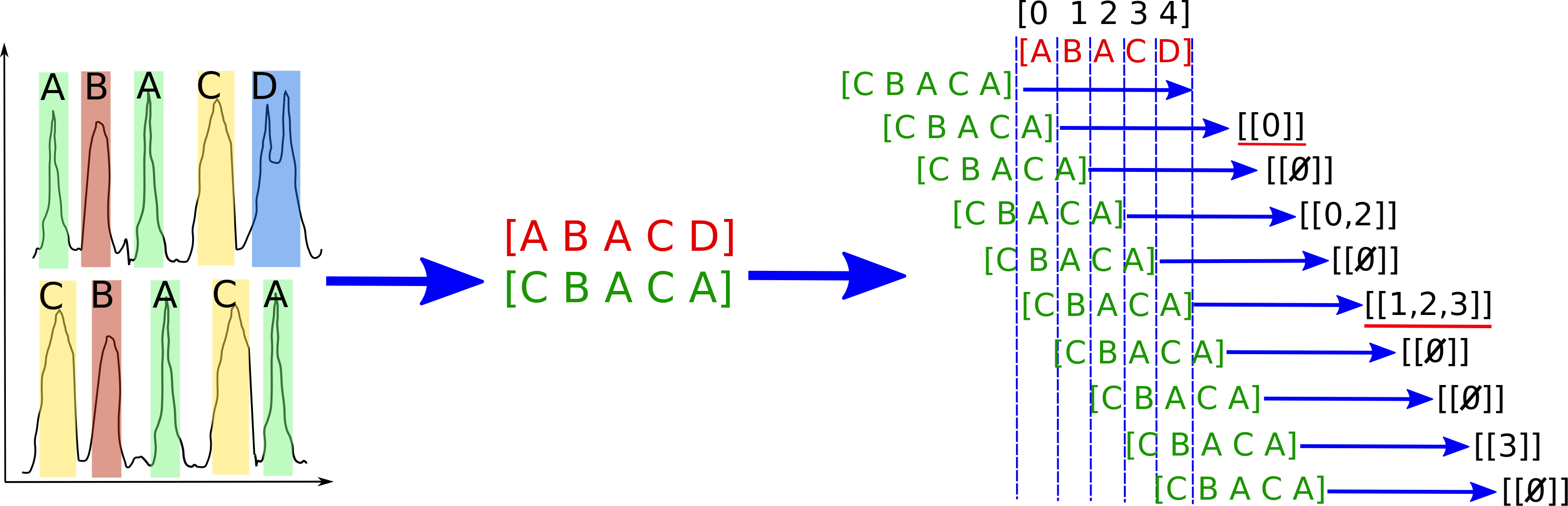}
    \caption{Schematic illustration of TS encoding and pulling procedure to produce subsets of set $V=\left\{ \left\{0 \right\}, \left\{ \emptyset \right\},\left\{ 0,2\right\},\left\{ \emptyset \right\} ,\left\{1,2,3 \right\},\left\{ \emptyset \right\},\left\{ \emptyset \right\},\left\{ 3\right\}, \left\{\emptyset\right\} \right\}$. The optimal selection to cover $U=\left\{ 0,1,2,3,4 \right\}$  in this case would be underlined subsets $V=\left\{ \left\{0 \right\}, \left\{1,2,3 \right\} \right\}$ with item numbers 0 and 4.  }
    \label{fig:pulling}
\end{figure}

There are many ways to discretize time series data, as reviewed previously.  For our purposes, we use the symbolic Fourier approximation (SFA) method \cite{schafer2012sfa}, as it provides differentiation between separate TS classes and features in high-dimensional data sets, allowing us to use these representative symbols for our quantum algorithm. Nevertheless, the exact discretization is data-dependent, with various hyperparameters (such as number of letters in the alphabet, length of each encoded string, etc.) present in the method; for a full explanation we refer the reader to~\cite{schafer2012sfa}. Given the encoded strings, we compare the time series using the following pulling procedure, illustrated in Figure~\ref{fig:pulling}. This pair-wise comparison is considered a preprocessing step necessary to formulate our set cover problem. Starting with one fixed string (red in the figure), we consider each encoded character as an independent element in the universe set\footnote{It is important to note that by using the same encoding scheme for all TS data, we ensure that all string characters belong to the same alphabet.} ($U=\{0,1,2,3,4 \}$ in the figure). 
A second string (green in the figure) is compared element-wise by successively moving the second string along the first, as illustrated. At every iteration, all character matches between the two strings are recorded as a new set. In the example from Figure~\ref{fig:pulling}, the set of sets is $V~=~\left\{ \left\{0 \right\}, \left\{\emptyset\right\},\left\{ 0,2\right\}, \left\{\emptyset\right\},\left\{1,2,3 \right\}, \left\{\emptyset\right\}, \left\{\emptyset\right\},\left\{ 3\right\}, \left\{\emptyset\right\} \right\}$.  \\

The procedure is repeated for the rest of the encoded training TS to form the set of sets $V$. This set, which is a union of all subsets obtained via the pulling technique, represents the features in common between the target time series and all other time series in the data set. Given this aggregate set, the goal is now to select the minimal subset that most closely reconstructs the universe, which is the NP-hard set cover problem. In the case illustrated in Figure~\ref{fig:pulling}, the optimal selection of subsets is underlined in red. In principle, solutions of this set cover problem do not preserve order of elements, and allow the use of the same element multiple times. This feature is useful for TS comparison, as elements of the time series data can be permuted and duplicated without affecting our reconstruction method. 

\subsection{Formulating the Set Cover Problem}
\label{sec:setcover}
Given the encoded strings and the common set of features $V$, we can now formulate the set cover problem as a QUBO, following the method demonstrated in~\cite{Lucas}. Consider the universe set $U=\left\{ 1, ..., n \right\}$, and a set of subsets $V_i$, such that $U=\bigcup\limits_{i=1}^{N} V_{i}$, $V_i \subseteq U$. Finding the smallest number of subsets $V_i$ whose union is $U$ is a well-known NP-hard optimization problem in the worst case~\cite{Karp}. In order to map the set cover problem to a QUBO problem, we use the following binary variables:
\begin{equation}
  x_i=\left\{
  \begin{array}{@{}ll@{}}
    1, & \text{if set }\ i \text{ is included},  \\
    0, & \text{otherwise,}
  \end{array}\right.
 \label{xi}
\end{equation} 
and 
\begin{equation}
  x_{\alpha,m}=\left\{
  \begin{array}{@{}ll@{}}
    1, & \text{if the number of}\ V_i \text{ which include element } \alpha \text{ is equal to } m,  \\
    0, & \text{otherwise.}
  \end{array}\right.
\label{xam}
\end{equation}
Here, $\alpha \in U$ denotes an element of universe set, and $m$ signifies if element $\alpha$ appears in $m$ subsets. We consider the full QUBO as a sum of two components: 
\begin{equation}
    H_A=A\sum_{\alpha=1}^n\left(1-\sum_{m=1}^N x_{\alpha,m}\right)^2+A\sum_{\alpha=1}^n\left(\sum_{m=1}^N mx_{\alpha,m}-\sum_{i:\alpha\in V_i}x_i\right)^2,
\label{HA}
\end{equation}
and
\begin{equation}
    H_B=B\sum_{i=1}^Nx_{i}.
\label{HB}
\end{equation} 
The complete QUBO is given by $H=H_A+H_B$ ~\cite{Lucas}. The first summation in $H_A$ imposes that exactly one of $x_{\alpha, m}$ must be selected in the minimum via a one-hot encoding. The second term in ${H_A}$ represents the number of times $\alpha$ is selected, and that this is equal to the number of selected subsets $\alpha$ appears in ($m$, as only one $x_{\alpha,m}$ can be 1 in the minimum). The final term $H_B$ (\ref{HB}) serves to minimize the number of $V_i$ needed to cover the universe $U$. The total number of variables required is $N+n(1 + M)$, where $M$ is the maximal number of sets that contain given element of $U$ (see~\cite{Lucas} for details). The limiting case where each element of $V_i$ included covers only one element of $U$ constrains the coefficient of $H_A$ and $H_B$ to $0<B<A$. The closer the coefficient $B$ and $A$, the more weight is given to (\ref{HB}), minimizing the number of elements selected from $V$. \\

\begin{figure}[h]
    \centering
     \includegraphics[totalheight=7cm]{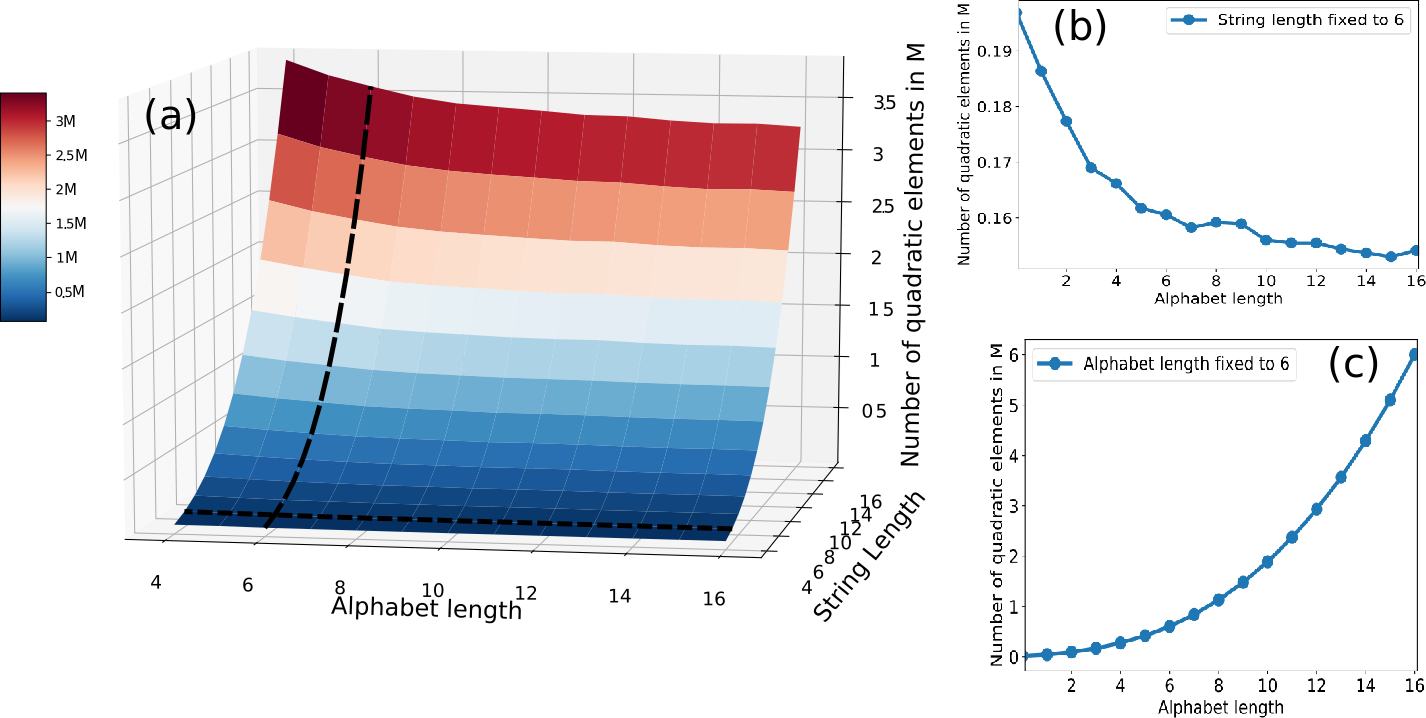}
    \caption{(a) The number of quadratic terms in millions as a function of string and alphabet length. (b) Quadratic elements as a function of alphabet length, with string length being fixed to 6. (c) Quadratic terms as function of string length, with alphabet length being fixed to 6. The corresponding isolines (b) and (c) are shown with dashed line on surface plot (a). }
    \label{fig:hamiltonianvars}
\end{figure}

In our application of time series reconstruction, the final size of the QUBO is heavily dependent on our choices during discretization. For example, the number of binary variables is equal to $N_{\text{train TS} }\left( 2L-1 \right)\left( L+1 \right)$, where $N_{\text{train TS} }$ is the number of TS in the training set used for reconstruction, and $L$ is the length of string that encodes the TS. Increasing the string length to encode each TS changes the size of the universe $U$. Allowing longer encoded strings to represent the data creates more subsets $V_i$. Therefore, there exists a trade-off between the granularity of the encoded strings and the ability to solve the set cover representation of the problem. Including more characters in our alphabet for discretization changes the non-empty sets $V_i$, which the number of quadratic elements in the QUBO depends on. The general trend is, however, that the number of the quadratic element decreases with the increase of the characters used in our alphabet. This could be explained due to the properties of the pulling procedure described above, since a smaller alphabet produces more non-empty elements $V_i$ which could be used for reconstruction of the universe $U$. In Figure~\ref{fig:hamiltonianvars} we show how varying these hyperparameters of the discretization affects the size of the QUBO problem, based on 20 test samples from the BeetleFly data set~\cite{hills2014classification}. \\

\section{Semi-supervised classification}
\label{sec:clustering}
We can now combine the methods described in the previous sections-- constructing the universe set $U$ from discretized data and the subsets $V$-- to perform semi-supervised clustering. We start by separating the input TS data into two groups-- training and test data. In our case we use training data sets with known labels, and the task we solve is to use the labeled data to assign labels to the test set. Normally, the training set with labeled data is significantly smaller than unlabeled test set, which we exploit in our method. \\ \newpage

We encode both the training and test data sets into strings using the method described in Section~\ref{sec:pulling}. We then perform the reconstruction procedure for every TS in our test set using the entire training set. Each TS from the test set is assumed to individually form a universe $U$, and is to be reconstructed using the sets $V_i$, obtained via the pulling procedure. Explicitly, using Figure~\ref{fig:pulling}, the red string is the TS from the test data set, and all strings in the training set are pulled through (greed strings) to obtain the $V_i$'s. This allows us to compare every test TS to the full training set in one-versus-all manner. Then, using the universe $U$ and $V_i$'s from the pulling procedure, we formulate the set cover problem outlined in Section~\ref{sec:setcover}. Thus, a single solution to that set cover problem (even sub-optimal in the worst case) allows us to reconstruct each TS from the test set using a set of discretized features obtained from \emph{all} elements which appear in the training set. Furthermore, since annealing-based sampling methods produce finite-temperature Boltzmann distributions~\cite{temperature}, various optima of the set cover problem could yield different ways to reconstruct the test TS using the training set. Due to this, it is therefore the users' task to use these reconstructions to associate each test TS with a label from the training set. We outline the steps of our classification procedure using pseudo-code in Algorithm~\ref{algo:semi_supervised}. \\

\begin{algorithm}
  \caption{Semi-supervised QUBO-based clustering algorithm }\label{algo:semi_supervised}
  \begin{algorithmic}
  \HEADER{Dimensionality reduction}
      \STATE Apply encoding of all TS into a string of characters using SFA~\cite{schafer2012sfa} or PAA~\cite{PAA_paper}-like methods. 
  \ENDHEADER
  \FOR{TS in test data}
    \HEADER{Produce set of sets for cover problem}
      \STATE Assume encoded TS forms the universe $U$. Apply puling procedure (Figure~\ref{fig:pulling}) to the full training data set to compute $V$ and map the problem to the set cover problem.
    \ENDHEADER
    \HEADER{Formulate the QUBO problem}
      \STATE Generate set cover QUBO an according to Section~\ref{sec:problem_formulation}.
    \ENDHEADER
    \HEADER{QUBO optimization}
      \STATE Find minimum of the QUBO; record subsets $V_i$ selected to cover $U$ .
    \ENDHEADER
    \HEADER{Analyze results of QUBO optimization}
      \STATE Trace back selected $V_i$ to respective time series from training set, and assign label to TS based on selection metrics.
    \ENDHEADER
  \ENDFOR
 \end{algorithmic}
\end{algorithm}


To classify the reconstructed test TS data we evaluated three different similarity metrics using set cover solutions: largest common subset $V_i$, highest number of common subsets $V_i$, and largest sum of common elements in selected $V_i$. We briefly explain how each metric is calculated, and discuss the performance of each.



\begin{itemize}
    \item \textbf{Largest common subset.} Given a candidate solution to the set cover problem, the label corresponding to the $V_i$ which contains the most elements is selected. The label is then assigned to the test TS. This metric captures the longest continuous set of features from the training TS data, and assumes that is sufficient to determine the label.
    \item \textbf{Number of common subsets.} Frequently, multiple $V_i$'s from the same training TS are used to reconstruct a test TS. In this metric, we count the number of $V_i$ subsets used to cover the universe. The test label is assigned the same label as the training TS which appears most frequently in the set cover solution.
    \item \textbf{Largest sum of subsets.} This metric is a combination of the previous two. For every training TS that is used to reconstruct a test set, the total number of elements used by each is counted (summed over all $V_i$'s). The label which corresponds to the training TS with the largest sum is assigned to the test TS. 
\end{itemize}


These metrics allow us to quantify the accuracy of our semi-supervised clustering algorithm. The first two metrics, being based on large sets of common features between the TS, performed the best (results shown in the next section). There was no significant difference between the two metrics, and the superiority of one metric over the other varied between data sets. The third metric, which was a combination of the first two, performed worse than either of the first metrics in the majority of the cases tested. While unexpected to begin with, this observation could be explained by the fact that because the third metric admits matches with many small subsets $V_i$ that are selected in the set cover, this metric could miss significant signatures present in the TS data. The largest common subset metric was selected for the experiments presented in the next section. It should also be noted that the use of labeled training data is not designed to not reach the accuracy of supervised learning methods. Moreover, there are modifications that could be made to the methods presented to improve the accuracy, for example increasing the word length and/or using a larger train set. Both are constrained in our use-case to prohibit excessively large QUBOs from being constructed. The goal of this method, as described, is to allow for relatively high accuracy using small sets of training data. \\

\begin{figure}[h!]
    \centering
     \includegraphics[scale=0.625]{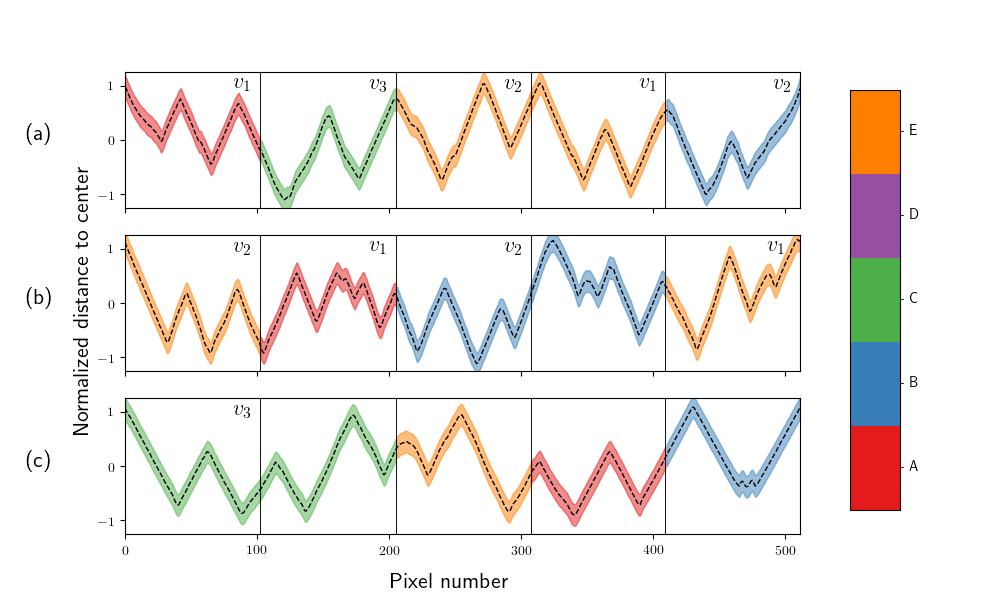}
    \caption{An illustrative example of reconstruction and classification from the BeetleFly data set. (a)~A test TS sample (encoded as `ACEEB') reconstructed from two training TS. Each box in the sub-figure is encoded as a single letter in a string, as per the color bar. The subsets $v_i$ obtained from the pulling procedure and used to reconstruct this data are shown both in the reconstructed (test) TS and in the training TS. (b)~The first training data used for reconstruction and classification (encoded as `EABBE'). (c)~A second time series used for reconstruction (encoded as `CCEAB'). }
    \label{fig:reconstruction_train_test}
\end{figure}

We provide an illustrative example of our QUBO-based reconstruction and classification in Figure~\ref{fig:reconstruction_train_test} using the BeetleFly data set. The task is to reconstruct the data in Figure~\ref{fig:reconstruction_train_test}~(a) using (b) and (c). For this example, an alphabet of size 5 was used for encoding, color-coded in the figure. The results of the set cover problem, formulated using the methods explained in previous sections, are three sets, shown as $v_1, v_2$, and $v_3$ in Figure~\ref{fig:reconstruction_train_test}. Meaning, each box (representing a fifth of the TS data per box) that appears in one of the subsets forming the solution is designated as such. Specifically, $v_1 = [$`A', `E'$]$, $v_2 = [$`E', `B'$]$, and $v_3 = [$`C'$].$ Therefore, the union $v_1 \bigcup v_2 \bigcup v_3 = U$, where $U = $`ACEEB', the test TS data to reconstruct. For classifying the reconstructed sample, we refer to the classes of the training data used for the reconstruction, and note that the training samples in Figure~\ref{fig:reconstruction_train_test}~(b) and (c) belong to two different classes. Using the similarity metrics defined above, it is easy to determine that $v_1$ and $v_2$ both originate from the time series (b), whereas only $v_3$ (which contains only a single element) is obtained from (c). Therefore, (a) is assigned the same label as (b). This example is representative of the majority of cases encountered during classification, with components of the reconstructed TS varying across multiple training samples, and often also across multiple classes. 

\section{Experiments and Results}
\label{sec:results}
\begin{figure}[h]
    \centering
     \includegraphics[totalheight=7cm]{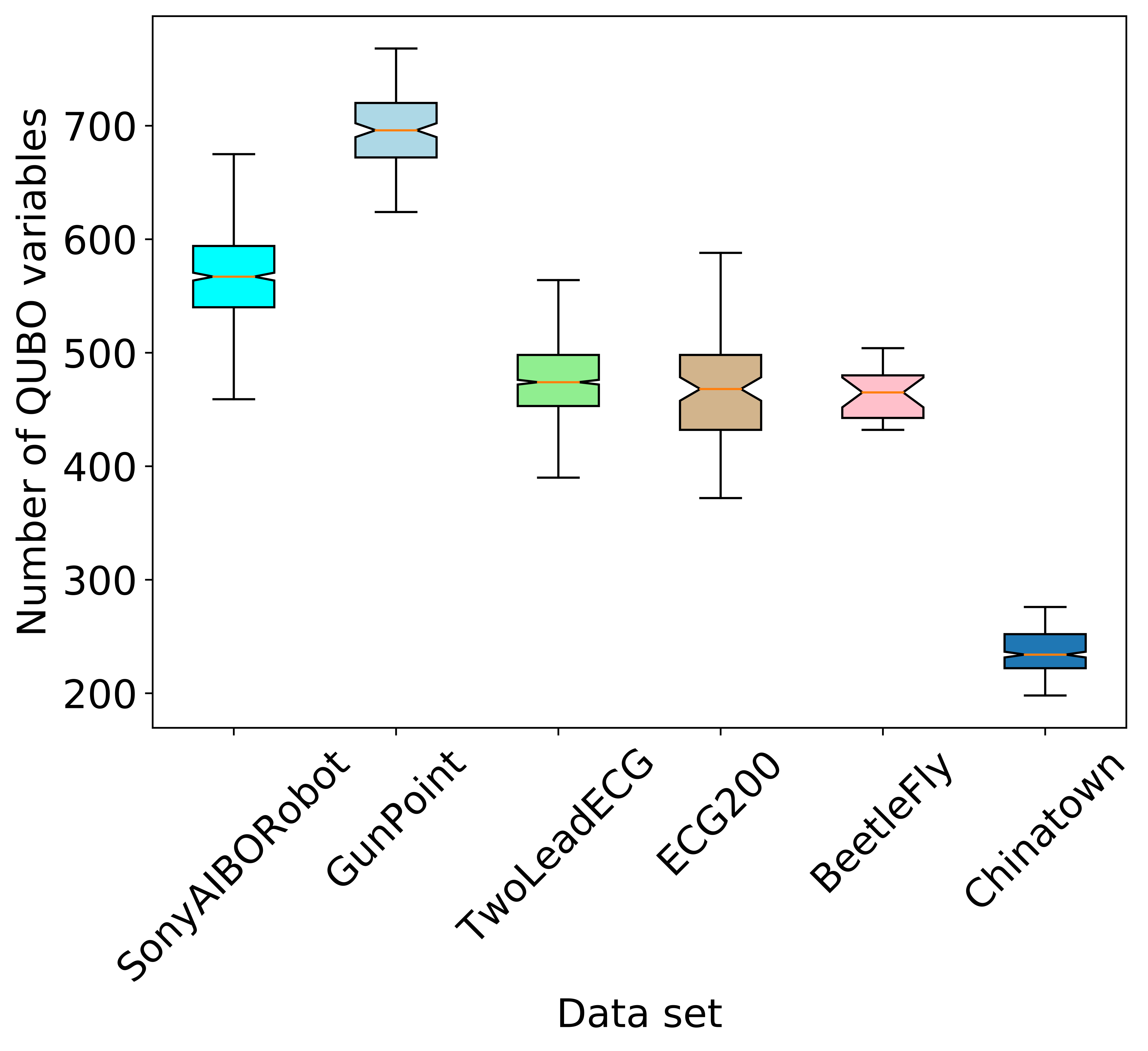}
    \caption{Distribution of number of QUBO variables for all data sets in Table \ref{table:dataparams}.}
    \label{fig:diag_distr}
\end{figure}

The experiments performed in this work used open-source labeled TS data available publicly~\cite{ts_class, ts_class_web}. We restricted our analysis to univariate TS data with two classes and small training set size to make our work amenable to NISQ devices in the near future. However, this method of semi-supervised classification can be used with any number of classes, at the cost of QUBO size. Since both the number of TS in the training data and the word length used to encode the TS contribute to the number of variables in the QUBO, we select data sets that have small numbers of TS in the training set. The exact sizes of the QUBOs for each data set are shown in Figure~\ref{fig:diag_distr}. The smallest QUBOs, from the Chinatown data set, were between 200-300 variables, and the largest were over 700 variables from the GunPoint data set. 

Due to their size, the QUBOs were optimized using simulated thermal annealing with 20,000 samples and 1000 sweeps~\cite{DWave_soft}, which was sufficient to ensure that low-energy local minima were sampled. The specific parameters used for the TS encoding are shown in Table~\ref{table:dataparams}. In general, the longer the TS are and the fewer TS are in the training set, the finer the discretization method required to accurately classify the test data. In all data sets we were able to reconstruct each test TS with elements from the training set, as explained in Section~\ref{sec:setcover}. To measure the performance of our classification method, we compared the accuracy of our labeling to semi-supervised and unsupervised classical classification methods, explained in more detail below. The results of these experiments for the various data sets are summarized in Table~\ref{table:dataresults}. \\

\begin{table}[ht]
\centering
\begin{tabular}{||c c c c c||} 
 \hline
 
 Data set & Data Type & \thead{Train/Test\\ Size } & \thead{Time series\\ Length} & \thead{Word/Alphabet\\ Length} \\
 \hline\hline
 SonyAIBORobotSurface1\cite{mueen2011logical}& Sensor& 10/601 & 70 &8/8 \\ 
 GunPoint\cite{conf/sdm/RatanamahatanaK05} & Motion & 30/150 & 150 & 5/5\\
 TwoLeadECG\cite{ECG_data} &ECG &20/1139 & 82 & 5/5 \\
 ECG200 \cite{ECG_data} &ECG &20/100 & 96 & 5/5  \\
BeetleFly\cite{hills2014classification} & Image&20/20 & 512 & 5/5 \\
 Chinatown \cite{China_data} & Traffic &20/345 & 24 & 5/5  \\  
 \hline
\end{tabular}

\caption{Table with data set description, number of TS in train and test sets, Length of TS, Length of each encoded string and number of different letters used to encode data set, clustering accuracy of measured on two classes and weighted average reported for QUBO-based and classical DTW-based methods. }
\label{table:dataparams}
\end{table}

\begin{table}[ht]
\centering
\begin{tabular}{||c c c c||} 
 \hline
 Data set & \thead{QUBO method \\ class1/class2/weighted }& \thead{K-means clustering\\ class1/class2/weighted  } &\thead{DTW semi-supervised\\ class1/class2/weighted  } \\
 \hline\hline
 SonyAIBORobotSurface1\cite{mueen2011logical}& 0.7/0.9/0.78 & 0.85/0.97/0.92& 0.97/0.63/0.83\\ 
 GunPoint\cite{conf/sdm/RatanamahatanaK05} & 0.76/0.79/0.78& 0.53/0.51/0.52& 0.82/0.77/0.79 \\
 TwoLeadECG\cite{ECG_data} & 0.6/0.62/0.61& 0.65/0.7/0.68 & 0.86/0.94/0.9  \\
 ECG200 \cite{ECG_data} & 0.61/0.82/0.75& 0.62/0.8/0.79 & 0.87/0.51/0.64  \\
BeetleFly\cite{hills2014classification} & 0.85/0.89/0.87 & 0.64/0.83/0.73 & 0.62/1.0/0.82 \\
 Chinatown \cite{China_data} & 0.72/0.91/0.86 & 0.37/0.78/0.67 &0.89/0.98/0.94 \\
 \hline
\end{tabular}

\caption{Table with data set description, number of TS in train and test sets, Length of TS, Length of each encoded string and number of different letters used to encode data set, clustering accuracy of measured on 2 classes and weighted average reported for QUBO-based and classical DTW-based methods. }
\label{table:dataresults}
\end{table}

As expected, the semi-supervised QUBO-based method outperforms classical unsupervised methods. To benchmark the results of our method, we use K-means clustering with pairwise DTW metrics calculated on the original TS (before encoding), with the labels being assigned based on belonging to one of two clusters. We also employ a classical semi-supervised method that is similar in spirit to the QUBO-based one, where the test TS labels are assigned by the DTW metric directly, calculated pairwise between each training and test TS (without encoding). We note however, that the QUBO-based method operates on a reduced dimensionality in contrast to the classical methods which use the original TS, where full information is preserved. Even under this consideration the accuracy of QUBO-based method is comparable with the semi-supervised DTW methods, and could be improved still by enriching the set $V$, i.e. by augmenting the training set or increasing the discretization granularity. \\

The worst performance of the QUBO-based algorithm is observed on the TwoLeadECG data set. This could be explained by the nature of our method, as well as the sensitivity of the ECG data. By using the set cover problem, we allow for permutations of subsets of TS data in the reconstruction of the test TS. It is likely that this permutation of TS segments, and similar representation in Fourier space of the signals from the two leads in the ECG measurements, makes our method not suitable for this kind of data. To confirm this is the case, and improve the classification accuracy, we applied SAX~\cite{senin2018grammarviz} encoding, based on sliding window time series magnitude, rather than the Fourier transform. Using this method, and encoding the TS as a word of length 5 constructed from a 5 letter alphabet, the accuracy is improved to 0.62 and 0.85 for the two respective classes (0.74 weighted average). In contrast to the Fourier encoding, the better results with ECG200 data are due to the significant differences between the classes of normal and ischemia ECG readings. \\

The highest accuracy is obtained using the BeetleFly and Chinatown data sets. In the first case, many permutations of the training set to construct the test set are permissible, which our method takes advantage of. The accuracy of our method is additionally improved by the relative size of the training set, further augmenting the combinatorial space of permutations. This robustness can also be explained by the dimensionality reduction technique for this data set: the 2D BeetleFly images (with different orientations) were mapped to 1D series of distances to the image centre, which again is beneficial for permutation-based methods. The Chinatown data set, for comparison, contained significantly shorter TS than BeetleFly. Encoding the Chinatown TS data with the same word length as BeetleFly resulted in higher granularity representations, and ultimately higher accuracy. This provides additional evidence that the accuracy of our method can be improved by increasing the granularity of the encoding. \\

\section{Conclusions}
\label{sec:discussion}
We present a QUBO-based method for TS reconstruction and semi-supervised classification that reaches accuracy scores comparable with classical DTW pairwise approaches, and in most cases outperforms unsupervised clustering. Among the advantages of our method is the utilization of significantly less data with respect to conventional classical methods, as well as a one-versus-all comparison that allows the selection of segments of data from multiple sources to reconstruct a single TS. This provides an additional robustness in our method in permutations of TS segments during the reconstruction. We showed how to reformulate the task of TS reconstruction as the set cover problem with a minimal number of subsets. In order to formulate this problem as a QUBO we apply TS dimensionality reduction by encoding each time series as a separate string. This encoding procedure and selection of comparison metrics (as discussed in Section~\ref{sec:clustering}) define the hyperparameter space of the problem. The QUBO-based classification method performed the best on image and traffic data, which is consistent with our method's inherit ability to utilize permutations of features/data within the TS to perform reconstruction.  \\

Time series reconstruction and classification has a wide variety of useful applications, such as: management of energy systems, factory process control, sensor systems, and many more. The methods introduced in this paper show how to reformulate the tasks of reconstruction and classification of such data using quantum computing. The fact that our work uses small training sets of labeled data means that the QUBOs produced could be solved by next-generation NISQ devices. Using quantum technologies, this method could analyze significantly more complex TS data, even in a live setting. The results of the optimization process (the selected subsets used for the reconstruction) would be informative as feedback for live process optimization as well. Future work in this are will be focused on generalising the method to multivariate TS cases, finding application-ready data sets, and execution of the presented methods on quantum processors. Specifically, with the advancement of hybrid quantum-classical algorithms, we will focus on converting the methods presented in this paper to be suitable for commercial applications.

\bibliographystyle{plain}
\bibliography{Manuscript.bib}
\end{document}